\documentclass[doublecol]{epl2}

\usepackage{color}
\usepackage{subfigure}
\usepackage{rotating}
\usepackage{graphicx}
\usepackage{epsf}
\usepackage{dcolumn}
\usepackage{bm}
\usepackage[mathlines]{lineno}

\usepackage{amssymb}
\usepackage{amsfonts}
\usepackage{amsmath}

\title{Electron-phonon coupling and superconductivity in \chem{LiB_{1+x}C_{1-x}}}

\author{Qi-Zhi Li\inst{1} \and Xun-Wang Yan\inst{2} \and Miao Gao\inst{1}\thanks{E-mail: \email{gaomiao@nbu.edu.cn}} \and Jun Wang\inst{1}\thanks{E-mail: \email{wangjun2@nbu.edu.cn}}}
\shortauthor{Q.-Z. Li \etal}

\institute{
  \inst{1} Department of Microelectronics Science and Engineering, Faculty of Science, Ningbo University - Zhejiang 315211, China\\
  \inst{2} College of Physics and Engineering, Qufu Normal University - Shandong 273165, China
}


\abstract{
By means of the first-principles density-functional theory calculation and Wannier interpolation, electron-phonon coupling and superconductivity are systematically explored for boron-doped LiBC (i.e. LiB$_{1+x}$C$_{1-x}$), with $x$ between 0.1 and 0.9. Hole doping introduced by boron atoms is treated through virtual-crystal approximation.
For the investigated doping concentrations, our calculations show
the optimal doping concentration corresponds to 0.8. By solving the anisotropic Eliashberg equations, we find that LiB$_{1.8}$C$_{0.2}$ is a two-gap superconductor, whose superconducting transition temperature, T$_c$, may exceed the experimentally observed value of MgB$_2$.
Similar to MgB$_2$, the two-dimensional bond-stretching $E_{2g}$ phonon modes along $\Gamma$-$A$ line have the largest contribution to electron-phonon coupling.
More importantly, we find that the first two acoustic phonon modes $B_1$ and $A_1$ around the midpoint of $K$-$\Gamma$ line play a vital role for the rise of T$_c$ in LiB$_{1.8}$C$_{0.2}$. The origin of strong couplings in $B_1$ and $A_1$ modes can be attributed to enhanced electron-phonon coupling matrix elements and softened phonons. It is revealed that all these phonon modes couple strongly with $\sigma$-bonding electronic states.}

\begin{document}

\maketitle

\section{Introduction}

The discovery of 39 K superconductivity in MgB$_2$ \cite{Nagamatsu-Nature_MgB2} has aroused great interest in searching for
new high-temperature superconductors whose pairing glue is electron-phonon coupling (EPC). Many compounds with similar atomic and/or
electronic structures to MgB$_2$ have been extensively investigated, such as $M$B$_2$ ($M$ = Be \cite{Satta-PRB64,Medvedeva-PRB64,Ravindran-PRB64}, Na \cite{Oguchi-JPSJ71}, Ca \cite{Medvedeva-PRB64,Ravindran-PRB64,Medvedeva-JETP,Choi-PRB80,Oguchi-JPSJ71}, Sc \cite{Medvedeva-PRB64,Oguchi-JPSJ71}, Cu \cite{Mehl-PRB64}, Sr
\cite{Ravindran-PRB64,Oguchi-JPSJ71}, Y \cite{Medvedeva-PRB64,Oguchi-JPSJ71}, Zr \cite{Oguchi-JPSJ71,Gasparov-JETP}, Ag \cite{Kwon-arXiv}, Ta \cite{Oguchi-JPSJ71,Rosner-PRB64}, Os \cite{Singh-PRB76}, and Au \cite{Kwon-arXiv}), CaBeSi \cite{Satta-PRB64}, LiBC \cite{Rosner-PRL_LiBC}, and MgB$_2$C$_2$ \cite{Ravindran-PRB64}.
Among these compounds, the most fascinating one is Li deficient Li$_x$BC, whose T$_c$ is predicted to be above 100 K for $x$ equal to 0.5.

In MgB$_2$, the underlying physics for high-T$_c$ superconductivity is the strong EPC between metallic covalent $\sigma$-bonding states and high-frequency $E_{2g}$ phonons associated with bond-stretching movements of boron atoms \cite{An-PRL86_4366,Y.Kong-PRB64_020501,Yildirim-PRL87_037001,Choi-PRB66_020513,Choi-Nature418_758}.
For semiconducting LiBC, the valence band maximum (VBM) locates at the $\Gamma$-$A$ line, corresponding to the $\sigma$-bonding states between boron and carbon atoms.
Rosner and coworkers suggested that the $\sigma$-bonding states can be rigidly lifted up to the Fermi level by removing some Li atoms \cite{Rosner-PRL_LiBC}, forming Li$_x$BC.
Thus the electronic structure of Li$_x$BC is reminiscent of that in MgB$_2$. And the high-T$_c$ superconductivity in Li$_x$BC seems natural. But evidence for superconductivity in Li$_x$BC is not available \cite{Bharathi-SSC124_423,Souptela-SSC125_17,Fogg-PRB67_245106}.
As a response to Li deficiency, the boron-carbon layer has drastic lattice distortions,
which diminish the hope to metallize the boron-carbon $\sigma$-bonding states \cite{Fogg-JACS128_10043}.

Considering the importance of Li atom in holding the crystal structure, replacing a certain amount of carbon atoms by boron atoms is regarded as a feasible way to realize hole doping in LiBC.
Miao \emph{et al.} used virtual-crystal approximation (VCA) to study the EPC of LiB$_{1.1}$C$_{0.9}$, and suggested the superconducting T$_c$ is about 36 K \cite{Miao-JAP113}.
We proposed a new compound Li$_3$B$_4$C$_2$ (i.e. LiB$_{1.33}$C$_{0.67}$), whose T$_c$ is about 53.8 K, based on
Wannier interpolation technique \cite{Gao-PRB91}. Another material Li$_4$B$_5$C$_3$ (LiB$_{1.25}$C$_{0.75}$), which is obtained through the substitution a BC$_3$ layer for one honeycomb B-C layer in LiBC,
is calculated to be superconducting under 16.8 K \cite{Bazhirov-PRB89}. Although the superconductivity in LiBC under several hole doping concentrations has been explored,
it is interesting to know how high the T$_c$ can reach and at which doping concentration the maximal T$_c$ can be obtained in LiB$_{1+x}$C$_{1-x}$ compounds.

In this work, we employ first-principles calculation and Wannier interpolation technique to investigate
the EPC and superconductivity in LiB$_{1+x}$C$_{1-x}$, with $x$ varying from 0.1 to 0.9 to determine the
optimal doping concentration. Our calculation shows that the highest T$_c$ can be achieved in LiB$_{1.8}$C$_{0.2}$. By solving the anisotropic Eliashberg equations,
it is found that LiB$_{1.8}$C$_{0.2}$ is a two-gap superconductor, whose T$_c$ may exceed the one of MgB$_2$ by a few Kelvin. At low doping, two-dimensional bond-stretching $E_{2g}$ phonon modes at $\Gamma$ point possess the
largest contribution to EPC. Further increasing the hole doping concentration,
$B_1$ and $A_1$ phonon modes at about $\frac{K\Gamma}{2}$ contribute enormously to the EPC, due to enhanced EPC matrix element and phonon softening.

\section{Methods}

In our calculations the plane wave basis method is used \cite{pwscf}.
We adopt the local density approximation (LDA) of Perdew-Zunger as the exchange-correlation functions.
The norm-conserving pseudopotentials \cite{Troullier-PRB43} are employed to model the electron-ion interactions.
Since boron and carbon are neighboring in the periodic table,
the hole doping introduced by boron is dealt with the VCA.
Hereafter the virtual atom is labeled as X, whose atomic mass is taken as the geometric mean value
for each doping level.
The kinetic energy cut-off and the charge density cut-off of the plane wave basis are chosen to be 80 Ry and 320 Ry, respectively.
The charge density is calculated on a 18$\times$18$\times$12 {\bf k}-point grid and a Methfessel-Paxton smearing \cite{Methfessel-PRB40} with width of 0.02 Ry.
For each doping concentration, the lattice constants and atomic positions are fully relaxed by minimizing the total energy [see Table I for details].
The phonons and the phonon perturbation potentials \cite{Giustino-PRB76} are calculated on a $\Gamma$-centered 6$\times$6$\times$4 mesh, within the framework of density-functional perturbation theory \cite{Baroni-RMP73_515}.

Maximally localized Wannier functions (MLWFs) \cite{Marzari-PRB56,Souza-PRB65} are constructed on a 6$\times$6$\times$4 grid of the Brillouin zone.
Here we use eight Wannier functions to describe the band structure of LiB$_{1+x}$C$_{1-x}$ around the Fermi level.
Two functions are $p_z$-like states associated with the X atoms, and six functions are $\sigma$-like
states localized in the middle of B-X bonds. For example, the spatial spreads
of the MLWFs that we generated in LiB$_{1.1}$C$_{0.9}$ are 2.79 {\AA}$^2$ for $p_z$-like states and 0.83 {\AA}$^2$ for $\sigma$-like states, respectively.
Fine electron (72$\times$72$\times$48) and phonon (24$\times$24$\times$16) grids are used to interpolate the electron-phonon coupling (EPC) quantities with Wannier90 \cite{Mostofi-CPC178,Mostofi-CPC185} and EPW codes \cite{Noffsinger-CPC181,Ponce-CPC209}.
Dirac $\delta$-functions for electrons and phonons are replaced by smearing functions with widths of 125 meV and 0.2 meV, respectively.
The EPC constant $\lambda$
can be determined through summation over the Brillouin zone
or integration of the Eliashberg spectral function $\alpha^2F(\omega)$ in frequency space as \cite{Allen-PRB6_2577,Allen-RPB12_905},
\begin{equation}
\label{eq:lambda}
\lambda=\frac{1}{N_q}\sum_{{\bf q}\nu}\lambda_{{\bf q}\nu}=2\int\frac{\alpha^2F(\omega)}{\omega}d\omega.
\end{equation}
The EPC constant $\lambda_{{\bf q}\nu}$ for mode $\nu$ at wavevector ${\bf q}$
is defined by \cite{Allen-PRB6_2577,Allen-RPB12_905},
\begin{equation}
\label{eq:lambda_qv}
\lambda_{{\bf q}\nu}=\frac{2}{\hbar N(0)N_k}\sum_{ij\bf k}\frac{1}{\omega_{{\bf q}\nu}}|g_{{\bf k},{\bf q}\nu}^{ij}|^2\delta(\epsilon^i_{\bf q})\delta(\epsilon^j_{\bf k+q}).
\end{equation}
Here $N_q$/$N_k$ is the total number of {\bf q}/{\bf k} points in the fine Brillouin-zone mesh. $N(0)$ is the electronic density of states (DOS) at the Fermi energy. ($i$, $j$) and $\nu$ denote indices of energy bands and phonon modes, respectively.
$\omega_{{\bf q}\nu}$ stands for the phonon frequency of the $\nu$-th phonon mode with wavevector ${\bf q}$.
$g_{{\bf k},{\bf q}\nu}^{ij}$ is the EPC matrix element.
$\epsilon_{\bf q}^i$ and $\epsilon_{\bf k+q}^j$ are
eigenvalues of Kohn-Sham states with respect to the Fermi energy at given bands and momentums.
The Eliashberg spectral function can be expressed as \cite{Allen-PRB6_2577,Allen-RPB12_905},
\begin{equation}
\label{eq:spectral}
\alpha^2F(\omega)=\frac{1}{2}\sum_{{\bf q}\nu}\delta(\omega-\omega_{{\bf q}\nu})\lambda_{{\bf q}\nu}\omega_{{\bf q}\nu}.
\end{equation}
Finally, the superconducting transition temperature (T$_c$) is determined by utilizing the McMillian-Allen-Dynes formula \cite{Allen-RPB12_905},
\begin{equation}
\label{eq:Tc}
\text{T}_c=\frac{\omega_{\text{log}}}{1.2}\exp\Big[\frac{-1.04(1+\lambda)}{\lambda(1-0.62\mu^*)-\mu^*}\Big].
\end{equation}
$\mu^*$ is the effective screened Coulomb repulsion constant, namely Coulomb pseudopotential. $\omega_{\text{log}}$
is the logarithmic average frequency, which can be computed through $\exp\Big[\frac{2}{\lambda}\int\frac{d\omega}{\omega}\alpha^2F(\omega)\log\omega\Big]$.

\section{Result and Analysis}

\begin{figure}[htb]
  \onefigure[width=8.6cm]{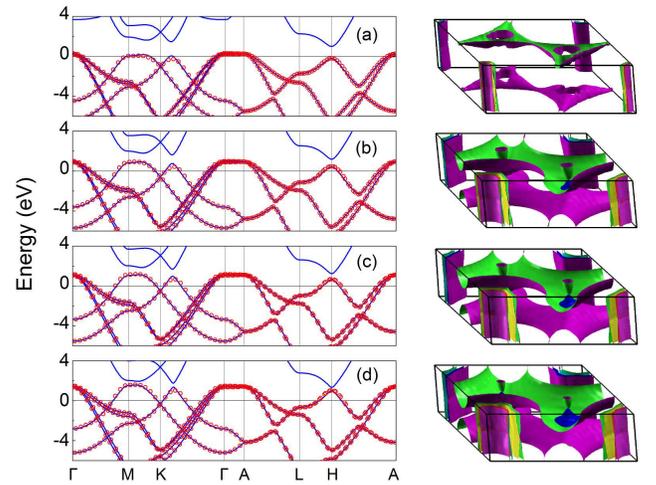}
  \caption{(Color on-line) The band structures and Fermi surfaces of LiB$_{1+x}$C$_{1-x}$, with $x$ being 0.1 (a), 0.5 (b), 0.6 (c), and 0.8 (d), respectively. The blue lines are calculated by the first principles, the red circles are obtained through MLWFs interpolation. The Fermi energy is set to zero. }
  \label{fig:bands}
\end{figure}

Figure~\ref{fig:bands} contains the band structures of LiB$_{1+x}$C$_{1-x}$.
As we see, LiB$_{1.1}$C$_{0.9}$ already becomes a metal. Namely, the $\sigma$-bonding states at the VBM
have been successfully metallized.
Thus the hole cylinders around $\Gamma$-$A$ line are from $\sigma$-bonding bands [right panel of Fig.~\ref{fig:bands}(a)].
Increasing hole doping concentration, the occupied energy bands almost move upward rigidly and the volumes enclosed by Fermi surfaces expand. While the location of empty energy bands
is not affected by doping. As a consequence, the energy gap is gradually reduced, and close to zero for LiB$_{1.8}$C$_{0.2}$.
The band structures obtained through interpolation of MLWFs are in excellent agreement with the ones calculated from first principles.
It is noted that the empty energy bands are not included in the Wannier interpolation and subsequent EPC calculation.
At $H$ point, the value of conduction band minimum (CBM) is at least 1.0 eV.
Physically, the electronic states that involved in EPC process are restricted to the Fermi level.
Thus the exclusion of empty energy bands does not affect the accuracy of our EPC results.

\begin{figure}[htb]
  \onefigure[width=8.6cm]{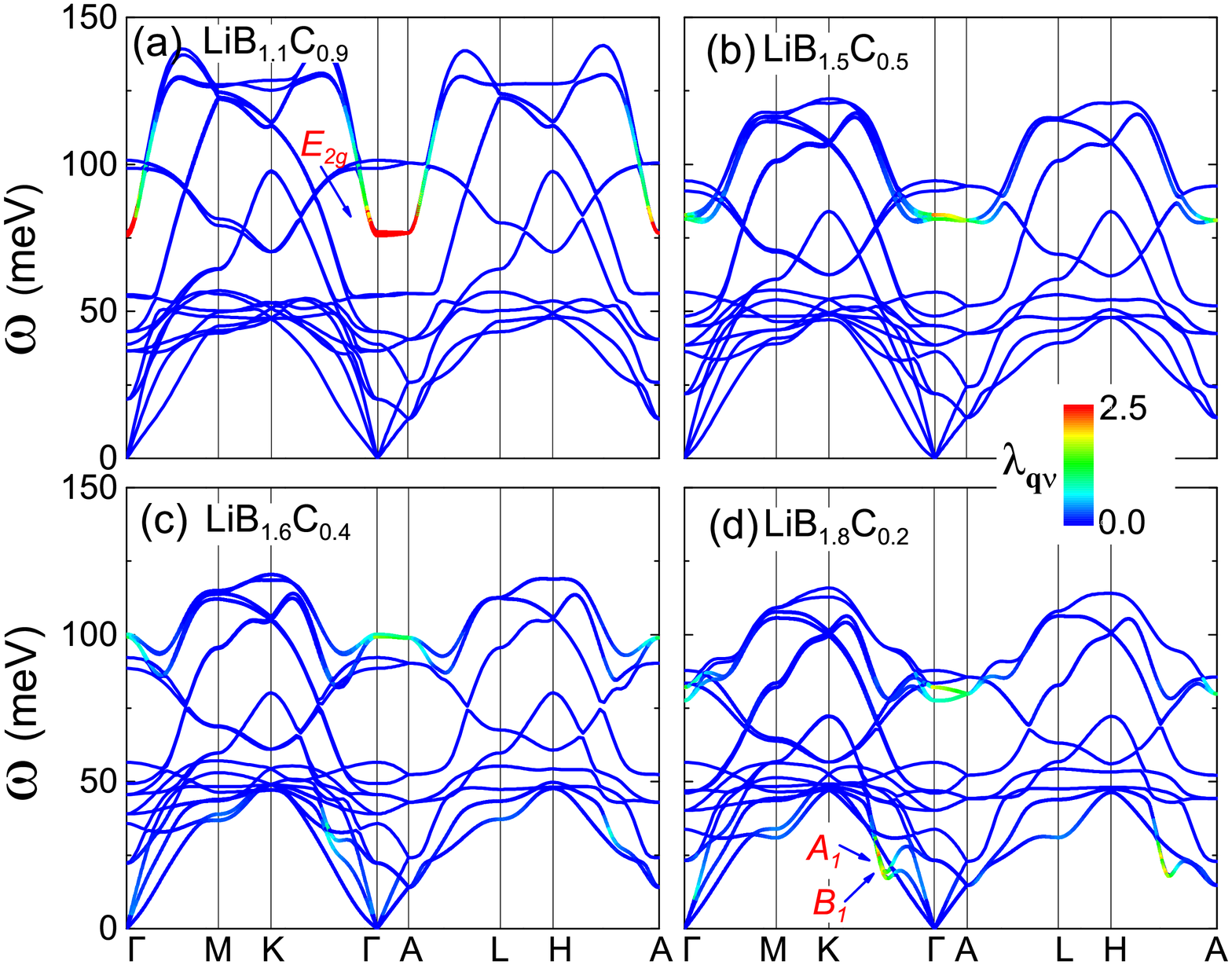}
  \onefigure[width=8.6cm]{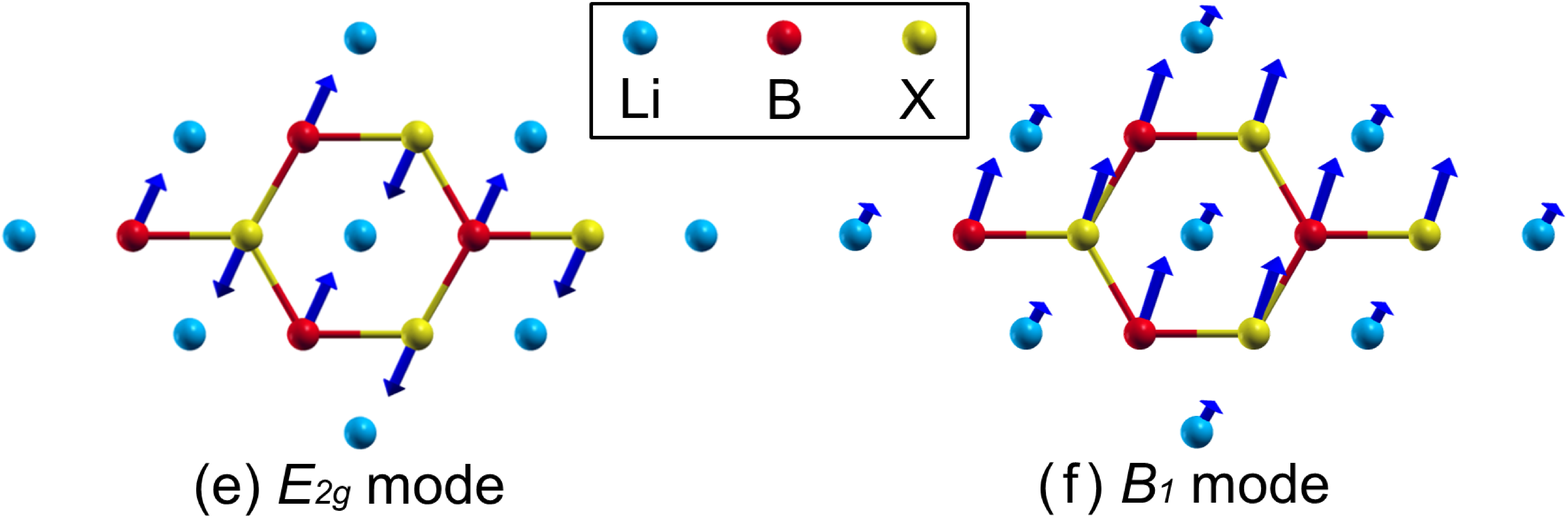}
  \caption{(Color on-line) The phonon spectra of LiB$_{1.1}$C$_{0.9}$ (a), LiB$_{1.5}$C$_{0.5}$ (b), LiB$_{1.6}$C$_{0.4}$ (c), and LiB$_{1.8}$C$_{0.2}$ (d).
  The wavevector {\bf q} and mode index $\nu$ resolved EPC constant $\lambda_{{\bf q}\nu}$ is color mapped.
  (e) and (f) are the top views of real-space vibrational patterns for $E_{2g}$ and $B_1$ phonon modes, respectively. The blue arrows and their lengths represent the directions and relative amplitudes of these vibration modes.}
  \label{fig:phonon-spectrum}
\end{figure}

The phonon spectra weighted by $\lambda_{{\bf q}\nu}$ are shown in Fig.~\ref{fig:phonon-spectrum}.
For the selected doping interval, no imaginary phonon frequency is found, indicating
the dynamical stability of LiB$_{1+x}$C$_{1-x}$. The strong EPC phonon modes, corresponding to the red lines along $\Gamma$-$A$,
are $E_{2g}$ modes, whose frequencies are listed in Table I.
Although the frequency of $E_{2g}$ mode fluctuates with the increase of doping concentration, an overall phonon softening is observed, especially, the first two acoustic phonon modes $B_1$ and $A_1$ at about $\frac{K\Gamma}{2}$ or $\frac{HA}{2}$ in LiB$_{1.8}$C$_{0.2}$ [see Fig.~\ref{fig:phonon-spectrum}(d)].
The vibrational configurations of $E_{2g}$ mode at $\Gamma$ and $B_1$ mode at $\frac{K\Gamma}{2}$
are shown schematically in Fig.~\ref{fig:phonon-spectrum}(e) and Fig.~\ref{fig:phonon-spectrum}(f), respectively.
The vibration of $A_1$ resembles that of $B_1$, but with even smaller Li displacement.
All these phonon modes mainly involve the two-dimensional movements of boron and virtual atoms.

\begin{figure}[t]
  \onefigure[width=8.6cm]{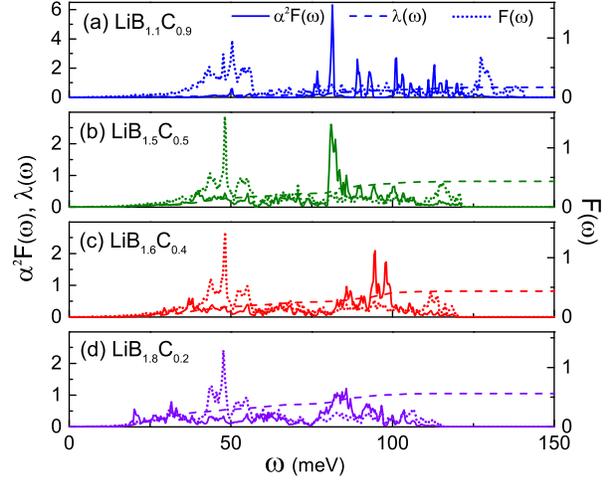}
  \caption{(Color on-line) The calculated Eliashberg spectral function $\alpha^2F(\omega)$, $\lambda(\omega)$, and phonon density of states $F(\omega)$ of LiB$_{1+x}$C$_{1-x}$ for $x$ equal to 0.1 (a), 0.5 (b), 0.6 (c), and 0.8 (d), respectively. $\lambda(\omega)$ is computed through $2\int_0^\omega \frac{1}{\omega'}\alpha^2F(\omega')d\omega'$. }
  \label{fig:a2f-lambda-omega}
\end{figure}
\begin{figure}[htb]
  \onefigure[width=8.6cm]{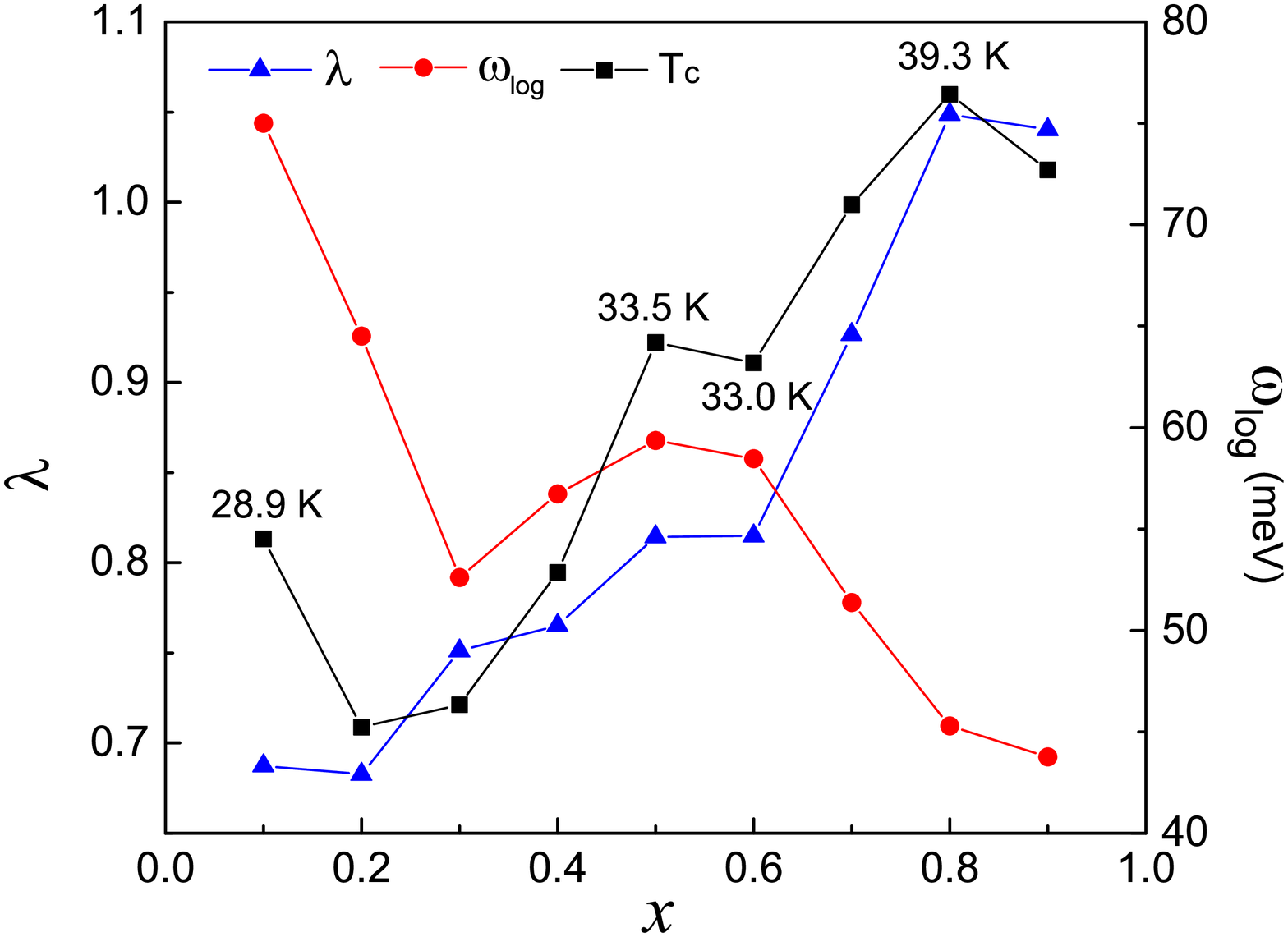}
  \caption{(Color on-line) The calculated EPC constant $\lambda$, logarithmic average frequency $\omega_{\text{log}}$, and superconducting T$_c$ for the investigated nine doping concentrations.
  The values of superconducting T$_c$ are tagged next to the data points.}
  \label{fig:lambda-omega-Tc}
\end{figure}

Compared with other three doping conditions, there is a very sharp peak for the Eliashberg spectral function $\alpha^2F(\omega)$ in LiB$_{1.1}$C$_{0.9}$ at about 80 meV associated with the strongly coupled $E_{2g}$ phonon modes [Fig.~\ref{fig:a2f-lambda-omega}(a)]. Nevertheless, the spectral weight from the low-frequency region is almost zero. The amplitude of $\alpha^2F(\omega)$ is gradually reduced with rising doping concentration, but the low-frequency part of $\alpha^2F(\omega)$ becomes stronger and stronger [Fig.~\ref{fig:a2f-lambda-omega}(b)-Fig.~\ref{fig:a2f-lambda-omega}(d)], consistent with the emergence of strong coupling $B_1$ and $A_1$ phonon modes in the phonon spectra [Fig.~\ref{fig:phonon-spectrum}].

\begin{table}[htb]
  \begin{center}
  \caption{The optimized lattice constants, $N(0)$ (states/eV /atom/spin), frequency of $E_{2g}$ (meV), $\lambda$, $\omega_{\text{log}}$ (meV) and T$_c$ (K) for different doping concentrations.}
  \begin{tabular}{ccccccc}
  \hline
  \hline
  $x$ & $a$ ({\AA}), $c/a$ & $N(0)$ & $\omega_{E_{2g}}$ & $\lambda$ & $\omega_{\text{log}}$  & T$_c$\\
  \hline
  0.1 & 2.750, 2.602 & 0.07 & 76.9 & 0.69 & 75.0 & 28.9 \\
  0.2 & 2.767, 2.587 & 0.11 & 86.2 & 0.68 & 64.5 & 24.5 \\
  0.3 & 2.786, 2.567 & 0.13 & 87.5 & 0.75 & 52.6 & 25.0 \\
  0.4 & 2.806, 2.547 & 0.13 & 80.5 & 0.77 & 56.7 & 28.1 \\
  0.5 & 2.825, 2.530 & 0.11 & 82.8 & 0.81 & 59.4 & 33.5 \\
  0.6 & 2.846, 2.507 & 0.11 & 99.3 & 0.81 & 58.5 & 33.0 \\
  0.7 & 2.871, 2.473 & 0.11 & 92.9 & 0.93 & 51.4 & 36.7 \\
  0.8 & 2.895, 2.446 & 0.11 & 82.1 & 1.05 & 45.3 & 39.3 \\
  0.9 & 2.919, 2.423 & 0.11 & 84.7 & 1.04 & 43.8 & 37.5 \\
  \hline \hline
  \end{tabular}
  \end{center}
\end{table}

The calculated EPC constant $\lambda$, logarithmic average frequency $\omega_{\text{log}}$, and T$_c$ are presented in Fig.~\ref{fig:lambda-omega-Tc}. For clarity, these quantities are also summarized in Table I. As we see, $\lambda$ and T$_c$ roughly display monotonous increasing versus the doping concentration.
LiB$_{1.8}$C$_{0.2}$ possesses the maximal superconducting T$_c$ of 39.3 K, slightly higher than that in MgB$_2$.
In the determination of T$_c$ for LiB$_{1+x}$C$_{1-x}$, the Coulomb pseudopotential $\mu^*$ is set to 0.1.
As pointed out in Ref.~\cite{Jin-PRB57}, an enhanced $N(0)$ can lead to an enhanced Coulomb pseudopotential.
For most doping concentrations, the $N(0)$ of LiB$_{1+x}$C$_{1-x}$ is around 0.11 states/eV/atom/spin [see Table I]. Thus the usage of a single $\mu^*$, i.e. 0.1, for
all the doping concentrations is reasonable. Interestingly, $N(0)$ is also equal to 0.11 states/eV/atom/spin in MgB$_2$ from our Wannier interpolation.

\begin{figure}[htb]
  \onefigure[width=8.6cm]{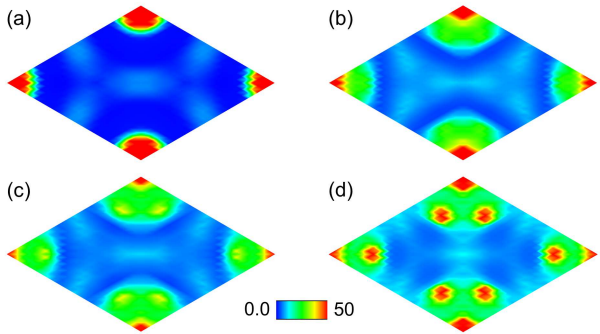}
  \caption{(Color on-line) Top views of $\lambda({\bf q}_{\text{2D}})$ in the reduced 2D Brillouin zone for LiB$_{1+x}$C$_{1-x}$ with $x$ being 0.1 (a), 0.5 (b), 0.6 (c), and 0.8 (d), respectively.}
  \label{fig:lambda_q}
\end{figure}

Since LiB$_{1.1}$C$_{0.9}$ has the largest $\lambda_{{\bf q}\nu}$ from $E_{2g}$ modes among these four doping situations along the $\Gamma$-$A$ line [Fig.~\ref{fig:phonon-spectrum}(a)],
it is surprising that the smallest $\lambda$ is found in LiB$_{1.1}$C$_{0.9}$.
This suggests that there exist sizeable $\lambda_{{\bf q}\nu}$s, which are not clearly reflected along the high-symmetry line in other three doping levels.
To unambiguously confirm above assumption, we plot $\lambda({\bf q}_{\text{2D}})$, defined by $\sum_{q_z\nu}\lambda_{{\bf q}\nu}$, in reduced two-dimensional (2D) Brillouin zone [Fig.~\ref{fig:lambda_q}].
Even though there is a big red spot around $\Gamma$ point in LiB$_{1.1}$C$_{0.9}$, the $\lambda({\bf q}_{\text{2D}})$ is close to zero in other area of the Brillouin zone [Fig.~\ref{fig:lambda_q}(a)].
In sharp contrast, considerable $\lambda({\bf q}_{\text{2D}})$ emerges beside $\Gamma$ point, especially for LiB$_{1.8}$C$_{0.2}$ [Fig.~\ref{fig:lambda_q}(d)],
in which six bright spots obviously appear at $\frac{K\Gamma}{2}$ and its equivalent points. These spots exactly come from the $B_1$ and $A_1$ phonon modes [Fig.~\ref{fig:phonon-spectrum}(d)], which thus have vital importance for the enhancement of $\lambda$ and T$_c$ upon the increasing of hole doping.

\begin{figure}[htb]
  \onefigure[width=8.6cm]{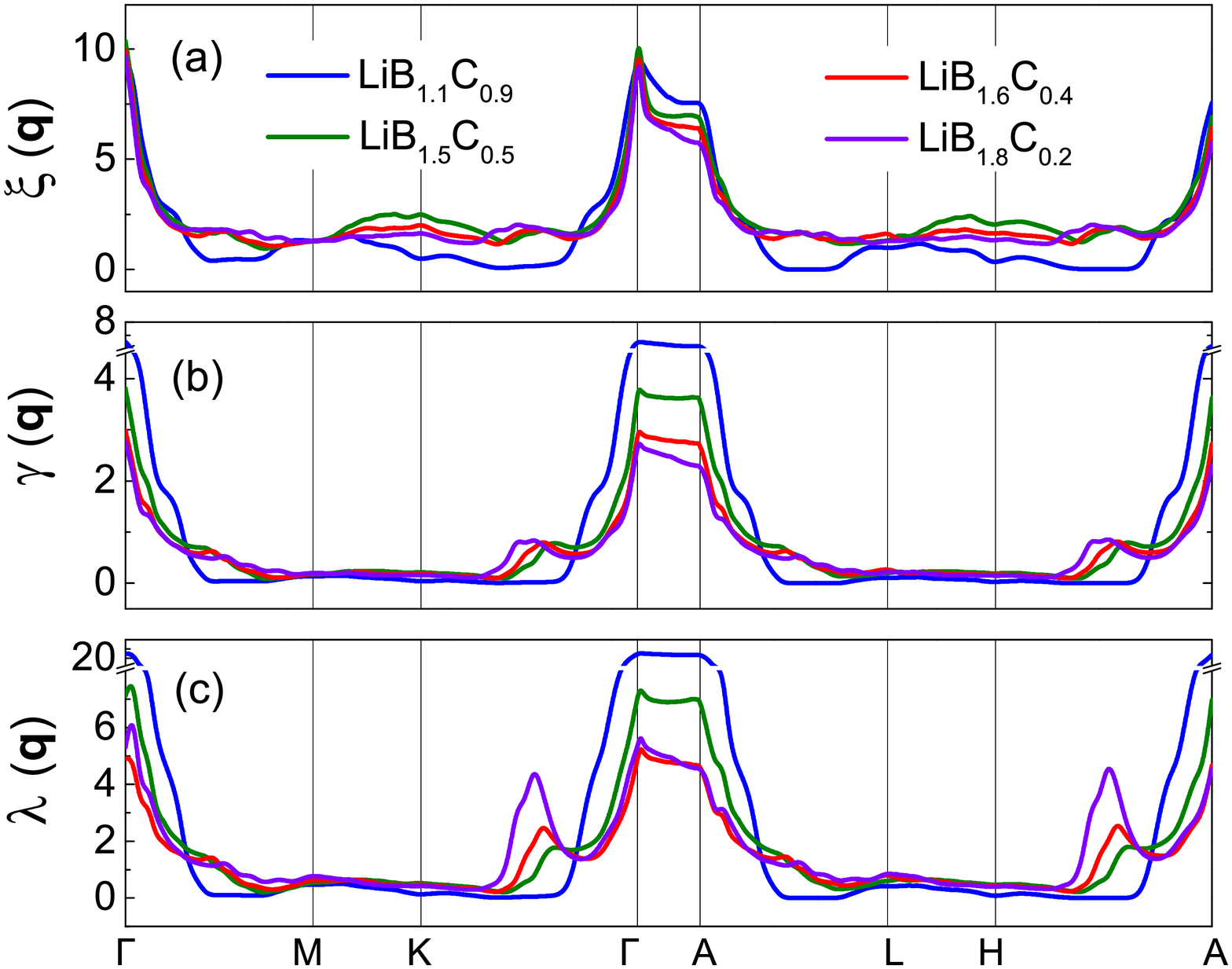}
  \caption{(Color on-line) $\xi({\bf q})$, $\gamma({\bf q})$, and $\lambda({\bf q})$ for LiB$_{1+x}$C$_{1-x}$.}
  \label{fig:nesting-gij}
\end{figure}

In order to identify the origin of the strong EPC $B_1$ and $A_1$ phonon modes, we calculate $\xi({\bf q})$ and $\gamma({\bf q})$, which read
\begin{equation}
\xi({\bf q})=\frac{1}{N_k}\sum_{ij\bf k}\delta(\epsilon_{\bf k}^i)\delta(\epsilon_{\bf k+q}^j)
\end{equation}
and
\begin{equation}
\gamma({\bf q})=\frac{1}{N_k}\sum_{ij\nu\bf k} |g_{{\bf k},{\bf q}\nu}^{ij}|^2\delta(\epsilon_{\bf q}^i)\delta(\epsilon_{\bf k+q}^j),
\end{equation}
respectively. $\xi({\bf q})$ is the Fermi surface nesting function. $\gamma({\bf q})$ is the summation of EPC matrix element $|g_{{\bf k},{\bf q}\nu}^{ij}|$ around the Fermi level. $\xi({\bf q})$ is almost the same with each other for the studied cases [Fig.~\ref{fig:nesting-gij}(a)]. This indicates that the Fermi surface nesting function $\xi({\bf q})$ is
not a dominant factor for strong EPC in $B_1$ and $A_1$ modes. We find that there are humps in $\gamma({\bf q})$ at the middle point of $K$-$\Gamma$ or $H$-$A$ for the later three concentrations [Fig.~\ref{fig:nesting-gij}(b)], reflecting the aggrandizement of EPC matrix elements around the Fermi level. These humps are further amplified in $\lambda(\bf q)$ by phonon softening [Fig.~\ref{fig:nesting-gij}(c) and Fig.~\ref{fig:phonon-spectrum}(d)]. Thus, both enhanced EPC matrix element $|g_{{\bf k},{\bf q}\nu}^{ij}|$ and softened $B_1$ and $A_1$ phonon modes account for the occurrence of strong EPC.

\begin{figure}[htb]
  \onefigure[width=8.6cm]{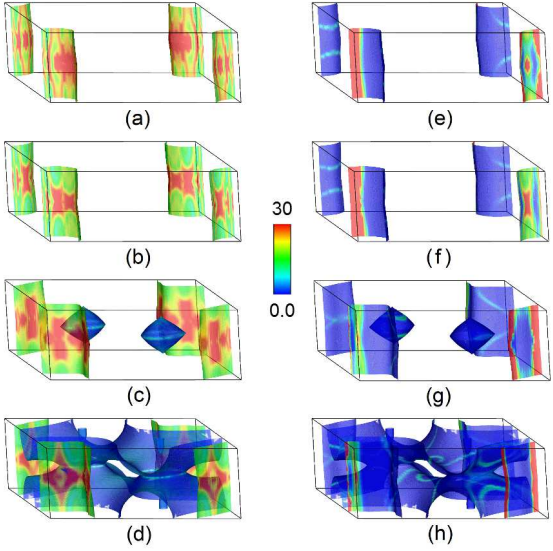}
  \caption{The calculated $\lambda_{{\bf k}i}$ in LiB$_{1.8}$C$_{0.2}$ for {\bf q} being $\Gamma$ (a-d) and $\frac{K\Gamma}{2}$ (e-h), respectively.
  There are four bands crossing the Fermi level in LiB$_{1.8}$C$_{0.2}$ [Fig.~\ref{fig:bands}(d)]. The Fermi surfaces formed by the four bands are plotted separately with ascending band energy from (a)/(e) to (d)/(h).}
  \label{fig:lambda_ki}
\end{figure}

Another important question should be addressed is
which electrons strongly couple with $E_{2g}$ mode at $\Gamma$ and $B_1$/$A_1$ mode at $\frac{K\Gamma}{2}$. Here we introduce a new quantity $\lambda_{{\bf k}i}$,
which represents the EPC constant at given momentum {\bf k} and band $i$.
\begin{equation}
\lambda_{{\bf k}i}=\frac{2}{\hbar N(0)}\sum_{j\nu}\frac{1}{\omega_{{\bf q}\nu}}|g_{{\bf k},{\bf q}\nu}^{ij}|^2\delta(\epsilon_{{\bf k}}^i)\delta(\epsilon_{{\bf k+q}}^j),
\end{equation}
Here $j$ is also the index of electronic energy band. By specifying the phonon wavevector ${\bf q}$, one can explicitly determine the electrons states that have large coupling with phonon modes at this wavevector. To be specific, we calculate $\lambda_{{\bf k}i}$ in LiB$_{1.8}$C$_{0.2}$ for {\bf q} being $\Gamma$ and $\frac{K\Gamma}{2}$, respectively [Fig.~\ref{fig:lambda_ki}].
It is found that the $E_{2g}$ phonon mode mainly couples with the cylinder-like hole Fermi surfaces around $\Gamma$-$A$ line [Fig.~\ref{fig:lambda_ki}(a)-Fig.~\ref{fig:lambda_ki}(d)]. This is also the case for $B_1$ and $A_1$ modes [Fig.~\ref{fig:lambda_ki}(e)-Fig.~\ref{fig:lambda_ki}(h)]. As we know, these cylindrical Fermi surfaces correspond to the B-X $\sigma$-bonding bands. So the high-T$_c$ superconductivity in LiB$_{1+x}$C$_{1-x}$ originates from strong coupling between phonon modes (i.e. $E_{2g}$ mode at $\Gamma$ and $B_1$/$A_1$ mode at $\frac{K\Gamma}{2}$) and B-X $\sigma$-boding bands.

With respect to the semi-empirical McMillian-Allen-Dynes formula, the superconducting T$_c$ can also be determined by solving the anisotropic Eliashberg
equations. Besides T$_c$, this method can provide more information about the superconducting gap structure.
For example, the two-gap structure in MgB$_2$ is clearly revealed by anisotropic Eliashberg calculations \cite{Margine-PRB87,Aperis-PRB92,Choi-PRB66,Choi-Nature418}.
Bekaert \emph{et al.,} further studied the novel superconducting gap resulted from the surface state in monolayer MgB$_2$ \cite{Bekaert-PRB96} or few-layer MgB$_2$ \cite{Bekaert-SciRep7}.

\begin{figure}[t]
  \onefigure[width=8.6cm]{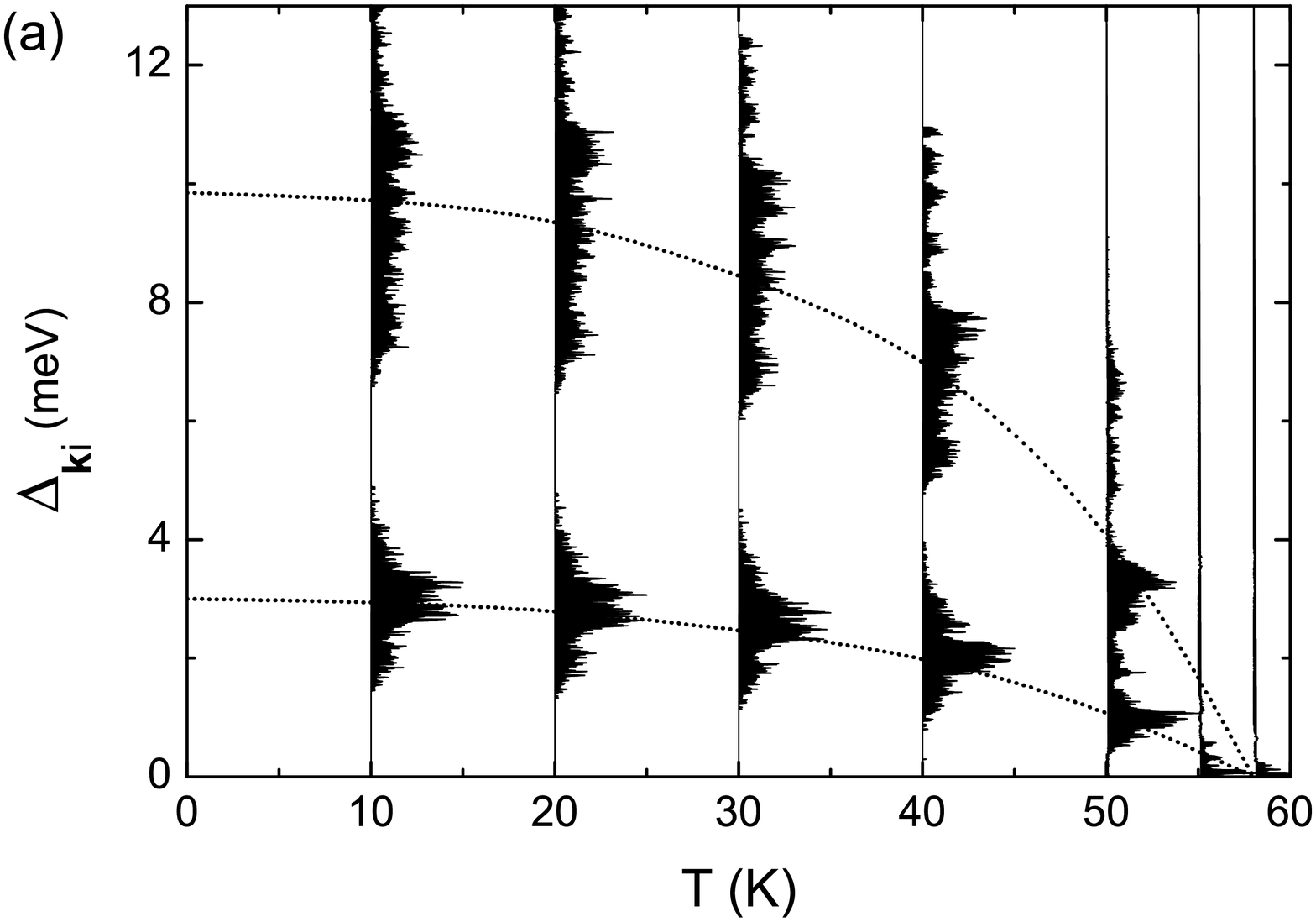}
  \onefigure[width=8.6cm]{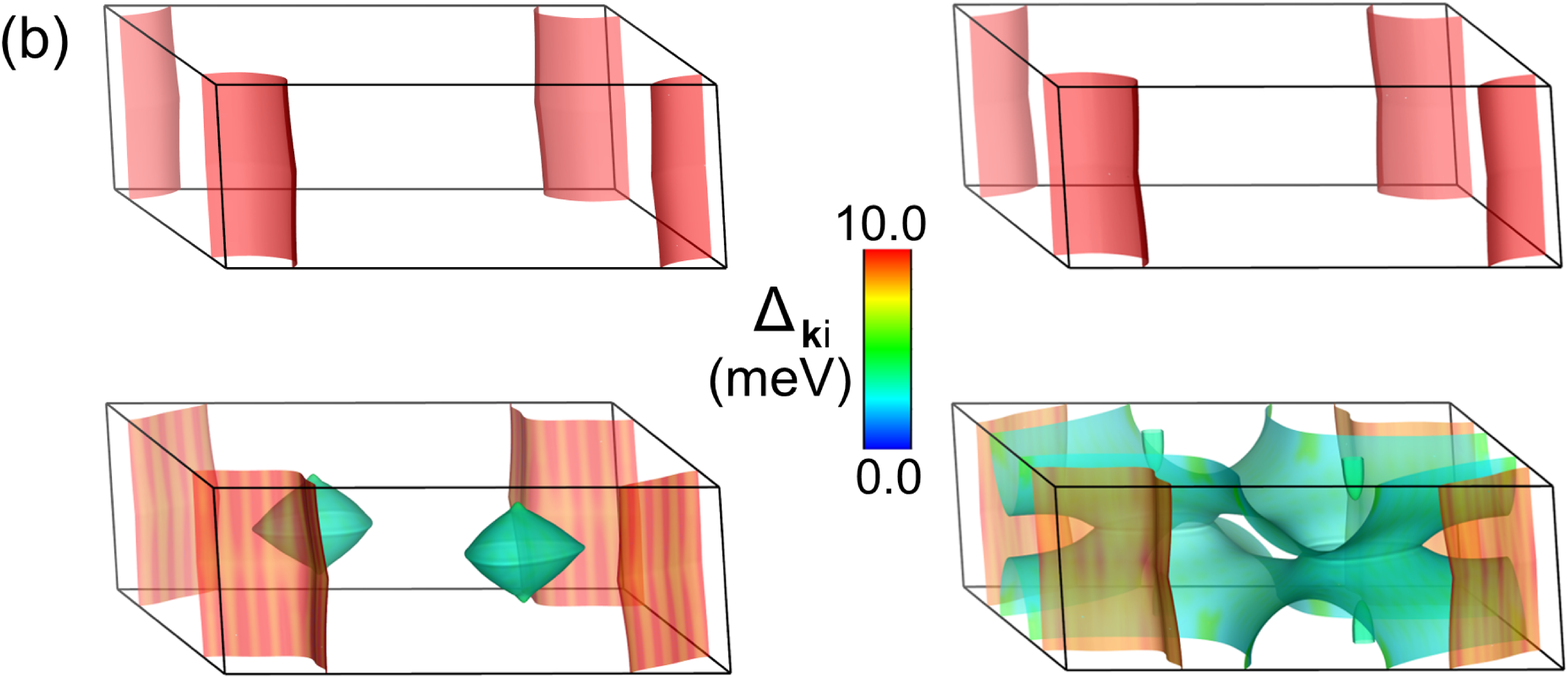}
  \caption{Superconducting properties of LiB$_{1.8}$C$_{0.2}$ obtained by solving the anisotropic Eliashberg equations.
  (a) Calculated anisotropic superconducting gap $\Delta_{{\bf k}i}$ on the Fermi surface as a function of temperature.
  The histograms at each temperature indicate the number of states on the Fermi surface with that superconducting gap energy. The dotted lines are guides to the eye.
  (b) Distribution of superconducting energy gaps on the four Fermi surfaces at 10 K.}
  \label{fig:Gap}
\end{figure}

For the sake of making T$_c$ in LiB$_{1+x}$C$_{1-x}$ more reasonable, we solve the anisotropic Eliashberg equations along the imaginary axis for LiB$_{1.8}$C$_{0.2}$.
The superconducting gaps are determined from the approximate Pad\'{e} continuation, and the Coulomb potential is chosen to be 0.16.
The reason for using 0.16 is to directly compare our result with that of MgB$_2$ given in Ref. \cite{Margine-PRB87,Ponce-CPC209}.
We can identify two distinct sets of superconducting gaps for LiB$_{1.8}$C$_{0.2}$ [Fig.~\ref{fig:Gap}(a)], which are associated
with the $\sigma$ and the $\pi$ sheets of the Fermi surface [Fig.~\ref{fig:Gap}(b)].
By taking the Fermi-surface averages, these gaps are $\Delta_\pi$=2.8 meV and $\Delta_\sigma$=9.5 meV at 10 K, which are about 12\% and 5.6\% larger than that in MgB$_2$ \cite{Ponce-CPC209}, respectively.
The two superconducting gaps vanish at 58 K, which is 7 K higher than the value of MgB$_2$ obtained by Ref. \cite{Ponce-CPC209} with the same calculation method.
It is noteworthy that anharmonic effect of phonons should be taken into consideration to reconcile theoretically calculated T$_c$ with experimentally observed one for MgB$_2$ \cite{Choi-PRB66,Choi-Nature418,Yildirim-PRL87}.
But the investigation of anharmonic effect in LiB$_{1+x}$C$_{1-x}$ is not the purpose of this work.
The hot zones of anisotropic electron-phonon coupling strength $\lambda$ are mainly distributed on the $\sigma$ Fermi sheets [Fig.~\ref{fig:Gap}(b)]. This further confirms the results presented in Fig.~\ref{fig:lambda_ki}.

\section{Discussion and Conclusion}
The sampling points contained in our fine {\bf k}-mesh for electrons and {\bf q}-mesh for phonons are 42 times more than that used in previous simulation of LiB$_{1.1}$C$_{0.9}$ \cite{Miao-JAP113}.
Thus our T$_c$ of LiB$_{1.1}$C$_{0.9}$ should be more reliable.
It is noted that
the superconducting T$_c$ of Li$_3$B$_4$C$_2$ \cite{Gao-PRB91} or Li$_4$B$_5$C$_3$ \cite{Bazhirov-PRB89} is different from the trend determined
in our calculations. The reasons for this inconsistency are twofold. Firstly, the distribution of boron and
carbon atoms in LiB$_{1+x}$C$_{1-x}$ is not included in the VCA. Different distributions will result in different electronic states, phonons, and T$_c$.
Secondly, the crystal structures used by Li$_3$B$_4$C$_2$ and Li$_4$B$_5$C$_3$
may be not the ground-state structures at corresponding stoichiometry. Considering several intelligent crystal structure prediction methods have been developed, such as random sampling method \cite{Pickard-JPCM}, particle-swarm optimization \cite{Wang-PRB82}, and evolutionary
technique \cite{Oganov-JCP124}, the ground-state structure of LiB$_{1.8}$C$_{0.2}$ is called for to verify our prediction.

\begin{figure}[htb]
  \onefigure[width=8.6cm]{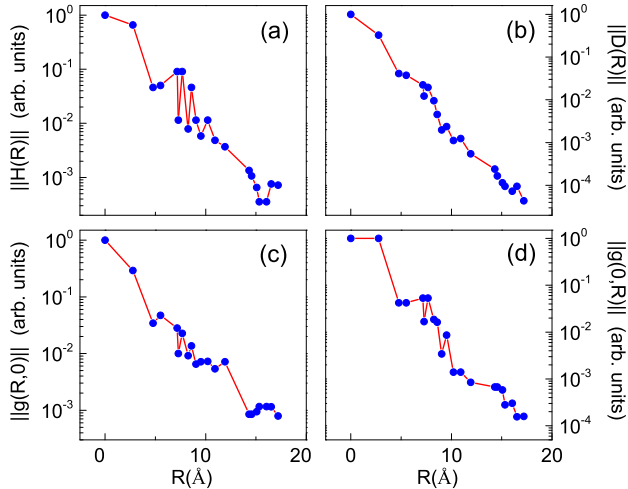}
  \caption{(Color on-line) Spatial decays of the electronic Hamiltonian $||$H(R)%
$||$, the dynamical matrix $||$D(R)$||$, and the EPC matrix element $||$%
g(R,0)$||$ and $||$g(0,R)$||$ in the Wannier representation for LiB$_{1.1}$C$_{0.9}$.
For the explicit definitions of above four quantities, please see Ref.~\cite{Giustino-PRB76}.}
  \label{fig:decay}
\end{figure}

In conclusion, we have extensively studied the EPC and phonon-mediated superconductivity for boron doped LiBC, utilizing the first-principles
calculations and the state-of-the-art Wannier interpolation technique.
At the VCA level, the maximal superconducting T$_c$ for LiB$_{1+x}$C$_{1-x}$ may slightly surpass that of MgB$_2$, with the optimal doping concentration $x$ being 0.8.
Beside the commonly known $E_{2g}$ phonon mode in MgB$_2$, we find that the $B_1$ and $A_1$ phonon modes, which mainly involve the
two-dimensional movements of boron and the boron-carbon virtual atoms, have strong coupling with the $\sigma$-bonding electronic states.

\acknowledgments

This research is supported by Zhejiang Provincial Natural Science Foundation of China under Grant No. LY17A040005,
and National Natural Science Foundation of China (Grant Nos. 11474004, 11404383, 11474174, and 11674185).
M.G. is also sponsored by K.C.Wong Magna Fund in Ningbo University.

\section{Appendix: Convergence test of EPC constant}

\begin{figure}[htb]
  \onefigure[width=8.6cm]{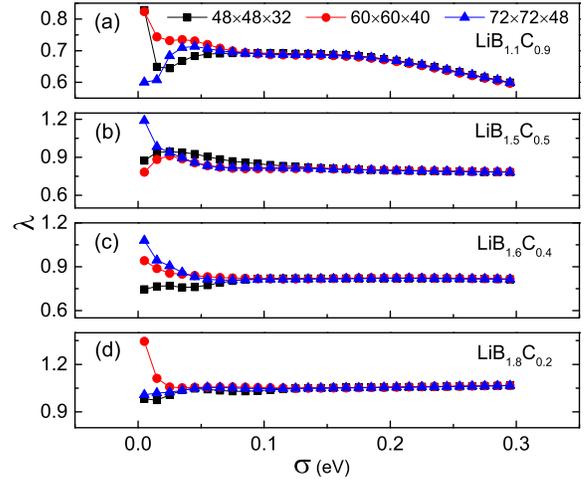}
  \caption{(Color on-line) Convergence test of EPC constant $\protect\lambda$
versus electron smearing $\sigma$. The fine {\bf q}-mesh is 24$\times$24$\times$16.
 (a) LiB$_{1.1}$C$_{0.9}$, (b)
LiB$_{1.5}$C$_{0.5}$, (c) LiB$_{1.6}$C$_{0.4}$, (d) LiB$_{1.8}$C$_{0.2}$. }
  \label{fig:convergence}
\end{figure}

The difficulty to obtain convergent EPC constant $\lambda$ stems from the so-called double-$\delta$-function approximation [Eq.~\eqref{eq:lambda_qv}].
Millions of {\bf k} points are commonly needed to accurately describe the electron-phonon scattering in the vicinity of the Fermi surface. This is time demanding and hardly accomplished by regular first-principles calculations. However, the state-of-the-art Wannier interpolation technique can achieve this goal by generalized Fourier transformations, simultaneously keeping the accuracy of the first principles \cite{Giustino-PRB76}.

Since the reliability of Wannier interpolation strongly depends on the localizations of the electronic Hamiltonian, the dynamical matrix, and the EPC matrix element in the Wannier representation, thus we have carefully examined above quantities, which demonstrate excellent exponential decay [Fig.~\ref{fig:decay}]. Three sets of fine {\bf k}-meshes (i.e. 48$\times$48$\times$32, 60$\times$60$\times$40, and 72$\times$72$\times$48) are used to check at which smearing $\sigma$ the $\lambda$ is convergent [Fig.~\ref{fig:convergence}].
As $\sigma$ approaching the zero limit, the curves of $\lambda$ calculated by the two denser {\bf k}-meshes begin to bifurcate for $\sigma$ being 125 meV [Fig.~\ref{fig:convergence}(c)]. Thus we regard the EPC properties gained at 125 meV as the convergent ones, which have been presented in the main text. It is noted that the convergence of {\bf k}-dependent superconducting gap $\Delta_{{\bf k}i}$ is more challenging than for $\lambda$. As shown by Margine \emph{et al.}, the energy distribution of $\sigma$ gap is changed from 2.5 meV to 1.5 meV when increasing the fine {\bf q}-mesh from 20$^3$ to 40$^3$, but the value of $\lambda$ is not affected \cite{Margine-PRB87}.
Here, careful examination for the convergence of $\Delta_{{\bf k}i}$ is not actualized, we adopt directly the convergent parameters for $\lambda$ to calculate the superconducting gap [Fig.~\ref{fig:Gap}].


\begin{thebibliography}{0}

\bibitem{Nagamatsu-Nature_MgB2}
  \Name{Nagamatsu J. \emph{et al.}}
  \REVIEW{Nature (London)}{410}{2001}{63}.

\bibitem{Satta-PRB64}
  \Name{Satta G., Profeta G., Bernardini F., Continenza A. \and Massidda S.}
  \REVIEW{Phys. Rev. B}{64}{2001}{104507}.

\bibitem{Medvedeva-PRB64}
  \Name{Medvedeva N. I., Ivanovskii A. L., Medvedeva J. E. \and Freeman A. J.}
  \REVIEW{Phys. Rev. B}{64}{2001}{020502(R)}.

\bibitem{Ravindran-PRB64}
  \Name{Ravindran P., Vajeeston P., Vidya R., Kjekshus A. \and Fjellv{\aa}g H.}
  \REVIEW{Phys. Rev. B}{64}{2001}{224509}.

\bibitem{Oguchi-JPSJ71}
  \Name{Oguchi T.}
  \REVIEW{J. Phys. Soc. Jpn.}{71}{2002}{1495}.

\bibitem{Medvedeva-JETP}
  \Name{Medvedeva N. I., Medvedeva J. E., Ivanovskii A. L., Zubkov V. G. \and Freeman A. J.}
  \REVIEW{JETP Lett.}{73}{2001}{336}.

\bibitem{Choi-PRB80}
  \Name{Choi H. J., Louie S. G. \and CohenM. L.}
  \REVIEW{Phys. Rev. B}{80}{2009}{064503}.

\bibitem{Mehl-PRB64}
  \Name{Mehl M. J., Papaconstantopoulos D. A. \and Singh D. J.}
  \REVIEW{Phys. Rev. B}{64}{2001}{140509(R)}.

\bibitem{Gasparov-JETP}
  \Name{Gasparov V. A., Sidorov N. S., Zver¡¯kova I. I. \and Kulakov M. P.}
  \REVIEW{JETP Lett.}{73}{2001}{532}.

\bibitem{Kwon-arXiv}
  \Name{Kwon S. K., Youn S. J., Kim K. S. \and Min B. I.}{ arXiv:cond-mat/0106483.}

\bibitem{Rosner-PRB64}
  \Name{Rosner H., Pickett W. E., Drechsler S.-L., Handstein A., Behr G., Fuchs G., Nenkov K., M¨¹ller K.-H. \and Eschrig H.}
  \REVIEW{Phys. Rev. B}{64}{2001}{144516}.

\bibitem{Singh-PRB76}
  \Name{Singh Y., Niazi A., Vannette M. D., Prozorov R. \and Johnston D. C.}
  \REVIEW{Phys. Rev. B}{76}{2007}{214510}.

\bibitem{Rosner-PRL_LiBC}
  \Name{Rosner H., Kitaigorodsky A. \and Pickett W. E.}
  \REVIEW{Phys. Rev. Lett.}{88}{2002}{127001}.

\bibitem{An-PRL86_4366}
  \Name{An J. M. \and Pickett W. E.}
  \REVIEW{Phys. Rev. Lett.}{86}{2001}{4366}.

\bibitem{Y.Kong-PRB64_020501}
  \Name{Kong Y., Dolgov O. V., Jepsen O. \and Andersen O. K.}
  \REVIEW{Phys. Rev. B}{64}{2001}{020501(R)}.

\bibitem{Yildirim-PRL87_037001}
  \Name{Yildirim T. \textit{et al.}}
  \REVIEW{Phys. Rev. Lett.}{87}{2001}{037001}.

\bibitem{Choi-PRB66_020513}
  \Name{Choi H. J., Roundy D., Sun H., Cohen M. L. \and Louie S. G.}
  \REVIEW{Phys. Rev. B}{66}{2002}{020513}.

\bibitem{Choi-Nature418_758}
  \Name{Choi H. J., Roundy D., Sun H., Cohen M. L. \and Louie S. G.}
  \REVIEW{Nature}{418}{2002}{758}.

\bibitem{Bharathi-SSC124_423}
  \Name{Bharathi A. \textit{et al.}}
  \REVIEW{Solid State Commun.}{124}{2002}{423}.

\bibitem{Souptela-SSC125_17}
  \Name{Souptela D., Hossainb Z., Behra G., L\"{o}sera W. \and Geibel C.}
  \REVIEW{Solid State Commun.}{125}{2003}{17}.

\bibitem{Fogg-PRB67_245106}
  \Name{Fogg A. M., Chalker P. R., Claridge J. B., Darling G. R. \and Rosseinsky M. J.}
  \REVIEW{Phys. Rev. B}{67}{2003}{245106}.

\bibitem{Fogg-JACS128_10043}
  \Name{Fogg A. M., Meldrum J., Darling G. R., Claridge J. B. \and Rosseinsky M. J.}
  \REVIEW{J. Am. Chem. Soc.}{128}{2006}{10043}.

\bibitem{Miao-JAP113}
  \Name{Miao R. \emph{et al.}}
  \REVIEW{J. Appl. Phys.}{113}{2013}{133910}.

\bibitem{Gao-PRB91}
  \Name{Gao M., Lu Z.-Y. \and Xiang T.}
  \REVIEW{Phys. Rev. B}{91}{2015}{045132}.

\bibitem{Bazhirov-PRB89}
  \Name{Bazhirov T., Sakai Y., Saito S. \and Cohen M. L.}
  \REVIEW{Phys. Rev. B}{89}{2014}{045136}.

\bibitem{pwscf}
  \Name{Giannozzi P. {\it et al.}}
  \REVIEW{J. Phys.: Condens. Matter}{21}{2009}{395502}.

\bibitem{Troullier-PRB43}
  \Name{Troullier N. \and Martins J. L.}
  \REVIEW{Phys. Rev. B}{43}{1991}{1993}.

\bibitem{Methfessel-PRB40}
  \Name{Methfessel M. \and Paxton A. T.}
  \REVIEW{Phys. Rev. B}{40}{1989}{3616}.

\bibitem{Giustino-PRB76}
  \Name{Giustino F., Cohen M. L. \and Louie S. G.}
  \REVIEW{Phys. Rev. B}{76}{2007}{165108}.

\bibitem{Baroni-RMP73_515}
  \Name{Baroni S., Gironcoli S. D., Corso A. D. \and Giannozzi P.}
  \REVIEW{Rev. Mod. Phys.}{73}{2001}{515}.

\bibitem{Marzari-PRB56}
  \Name{Marzari N. \and Vanderbilt D.}
  \REVIEW{Phys. Rev. B}{56}{1997}{12847}.

\bibitem{Souza-PRB65}
  \Name{Souza I., Marzari N. \and Vanderbilt D.}
  \REVIEW{Phys. Rev. B}{65}{2001}{035109}.

\bibitem{Mostofi-CPC178}
  \Name{Mostofi A. A., Yates J. R., Lee Y.-S., Souza I., Vanderbilt D. \and Marzari N.}
  \REVIEW{Comput. Phys. Commun.}{178}{2008}{685}.

\bibitem{Mostofi-CPC185}
  \Name{Mostofi A. A., Yates J. R., Pizzi G., Lee Y.-S., Souza I., Vanderbilt D. \and Marzari N.}
  \REVIEW{Comput. Phys. Commun.}{185}{2014}{2309}.

\bibitem{Noffsinger-CPC181}
  \Name{Noffsinger J., Giustino F., Malone B. D., Park C.-H., Louie S. G. \and Cohen M. L.}
  \REVIEW{Comput. Phys. Commun.}{181}{2010}{2140}.

\bibitem{Ponce-CPC209}
  \Name{Ponc\'{e} S., Margine E. R., Verdi C. \and Giustino F.}
  \REVIEW{Comput. Phys. Commun.}{209}{2016}{116}.

\bibitem{Allen-PRB6_2577}
  \Name{Allen P. B.}
  \REVIEW{Phys. Rev. B}{6}{1972}{2577}.

\bibitem{Allen-RPB12_905}
  \Name{Allen P. B. \and Dynes R. C.}
  \REVIEW{Phys. Rev. B}{12}{1975}{905}.

\bibitem{Jin-PRB57}
  \Name{Jin Y.-G. \and Chang K. J.}
  \REVIEW{Phys. Rev. B}{57}{1998}{14684}.

\bibitem{Margine-PRB87}
  \Name{Margine E. R.\and Giustino F.}
  \REVIEW{Phys. Rev. B}{87}{2013}{024505}.

\bibitem{Aperis-PRB92}
  \Name{Aperis A., Maldonado P., \and Oppeneer P. M.}
  \REVIEW{Phys. Rev. B}{92}{2015}{054516}.

\bibitem{Choi-PRB66}
  \Name{Choi H. J., Roundy D., Sun H., Cohen M. L., \and Louie S. G.}
  \REVIEW{Phys. Rev. B}{66}{2002}{020513(R)}.

\bibitem{Choi-Nature418}
  \Name{Choi H. J., Roundy D., Sun H., Cohen M. L., \and Louie S. G.}
  \REVIEW{Nature}{418}{2002}{758}.

\bibitem{Bekaert-PRB96}
  \Name{Bekaert J., Aperis A., Partoens B., Oppeneer P. M., \and Milo\v{s}evi\'{c} M. V.,}
  \REVIEW{Phys. Rev. B}{96}{2017}{094510}.

\bibitem{Bekaert-SciRep7}
  \Name{Bekaert J. {\it et al.}}
  \REVIEW{Sci. Rep.}{7}{2017}{14458}.

\bibitem{Yildirim-PRL87}
  \Name{Yildirim T. {\it et al.}}
  \REVIEW{Phys. Rev. Lett.}{87}{2001}{037001}.

  \bibitem{Pickard-JPCM}
  \Name{Pickard C. J. \and Needs R.}
  \REVIEW{J. Phys.: Condens. Matter}{23}{2011}{053201}.

\bibitem{Wang-PRB82}
  \Name{Wang Y., Lv J., Zhu L. \and Ma Y.}
  \REVIEW{Phys. Rev. B}{82}{2010}{094116}.

\bibitem{Oganov-JCP124}
  \Name{Oganov A. R. \and Glass C. W.}
  \REVIEW{J. Chem. Phys.}{124}{2006}{244704}.


\end{thebibliography}
\end{document}